\documentclass[11pt]{article}
\pdfoutput=1
\usepackage{amsmath}
\usepackage{graphicx}

\begin{document}

\begin{center}

\Large

{\bf 10 key problems in stellar dynamics: in retrospect}

\vspace{0.2in}

\large
(10 problems solution of which can seriously influence the stellar dynamics)

\vspace{0.3in}

Vahe G. GURZADYAN

\vspace{0.1in}

Department of Theoretical Physics, Yerevan Physics Institute, Armenia

\end{center}

\vspace{0.2in}

(Published in {\it ``Ergodic Concepts in Stellar Dynamics"}, Eds. V.G. Gurzadyan, D. Pfenniger, Lecture Notes in Physics, vol.430, p.281-284 (Problems), p.285-291 (Comments on Key Problems), Springer-Verlag, Berlin, 1994. See also the more recent: V.G. Gurzadyan, A.A. Kocharyan,
A \& A, 505, 625, 2009, arXiv:0905.0517.)

\vspace{0.5in}

{\it Theory}

\vspace{0.1in}

{\bf Problem 1.}

Creation of a  mathematical  model of $N$--body gravitating  system, 

enabling to either avoid compactness, measure and other difficulties 

or to define statistical properties (mixing, etc.)  without  use  of 

those conditions.

More concrete aim can be formulated as follows:

{\it Let $(X, {\cal B}(X), \mu, f)$ be a smooth dynamical system with
continuous or discrete time, where $X$ is not compact and/or
$\mu(x)=\infty$. The problem is to define: 

a.property of mixing;

b.correlation functions.}

\vspace{0.2in}

{\bf Problem 2.}

Study of behavior of time correlation functions for  physical  phase 
functions (kinetic energy, etc).

Strict formulation reads:

{\it Consider a dynamical system when $\forall g_1,g_2\in L^2(X)  
\hspace{0.1in} \exists C_{g_1,g_2}>0$ and $\beta_{g_1,g_2}>0$ such 
that $\forall t>0$ one has
$$
\mid b_{g_1,g_2}(t)\mid \leq C_{g_1,g_2}\exp{(-\beta_{g_1,g_2}t)},
$$
for $t\in R$ or $t\in Z$.

The problem then comes to:

a. finding out conditions the dynamical system with such properties 
should fulfill;

b. estimation of $\beta$.}

\vspace{0.2in}

{\bf Problem 3.}

Derivation  of  physical  conditions  to  describe  core  collapse, 

evaporation and other evolutionary effects of $N$--body systems.

\vspace{0.2in}

{\bf Problem 4.}

Study of the role of  stochasticity  and  regularity  of  motion  in 

determination of morphology of galaxies.

\vspace{0.2in}

{\it Computer Simulations}

\vspace{0.1in}

{\bf Problem 5.}

Creation of a computer code to describe the $N$--body system with 
phase trajectory close to that of the physical one for long enough
time scales;  the  same for systems with non-point particles.

{\it Let $x(t)$ be an exact solution of $N$--body problem, $x_c(t)$,
that of the calculated by computer by means of some method, and
$$
\varepsilon_c(T)=\sup_{t\in[0,T]} \|x_c(t)-x(t)\|.
$$
The problem is to:

a. evaluate $\varepsilon_c(T)$;

b. study the limit
$$
\varepsilon_c\equiv \lim_{t\to \infty} \sup \frac{\ln \varepsilon_c(T)}{T},
$$
and its relation to Lyapunov characteristic exponents;

c. find out methods for which $\varepsilon _c=0$.}

\vspace{0.2in}

{\bf Problem 6.}

Development of effective methods of numerical study  of  statistical 

properties of $N$--body systems, particularly  of  local  (in  time) 

characteristics of instability.

\vspace{0.2in}

{\bf Problem 7.}

Search of computer algebraic methods  of  study of  evolution  of 
gravitating systems, i.e. avoiding  the  numerical integration  of 
 differential  equations (iterations).

\vspace{0.2in}

{\it Observations}

\vspace{0.1in}

{\bf Problem 8.}

Increase of high accuracy data on central regions  of  galaxies  and 

star clusters,  including  the  run  by  radius  of number  density 

of stars, velocity dispersion, eccentricity of system, etc.

\vspace{0.2in}

{\bf Problem 9.}

Formulation  of  quantitative  empirical  relations  determining the 

position of the stellar system on the path of evolution.

\vspace{0.2in}

{\bf Problem 10.}

Search of empirical relations enabling to distinguish  the  role  of 

binary  and  $N$--body  gravitational  interactions  of  stars  in 

relaxation driving effects in galaxies and star clusters.

\newpage

{
\noindent \Large \bf Comments on "10 Key Problems"
}

\vspace{0.5in}
\textbf{Richard H. MILLER} 

\vspace{0.1in}
\textbf{1. General Remarks}
\vspace{0.1in}

Exponential separation of most (initially) neighboring trajectories in phase space is generally accepted in this audience, and coping with it is generally regarded as important in our understanding of stellar systems. However, we seem on several occasions to have become obsessed with the details at the expense of the broader picture.

The goal of studies in stellar dynamics is to understand the dynamics of real stellar systems-galaxies or star clusters. Logical connections in our understanding and interpretations of the dynamics of galaxies are shown in Fig. 1.

\begin{figure}[h]
\begin{center} 
\includegraphics[width=0.65\textwidth]{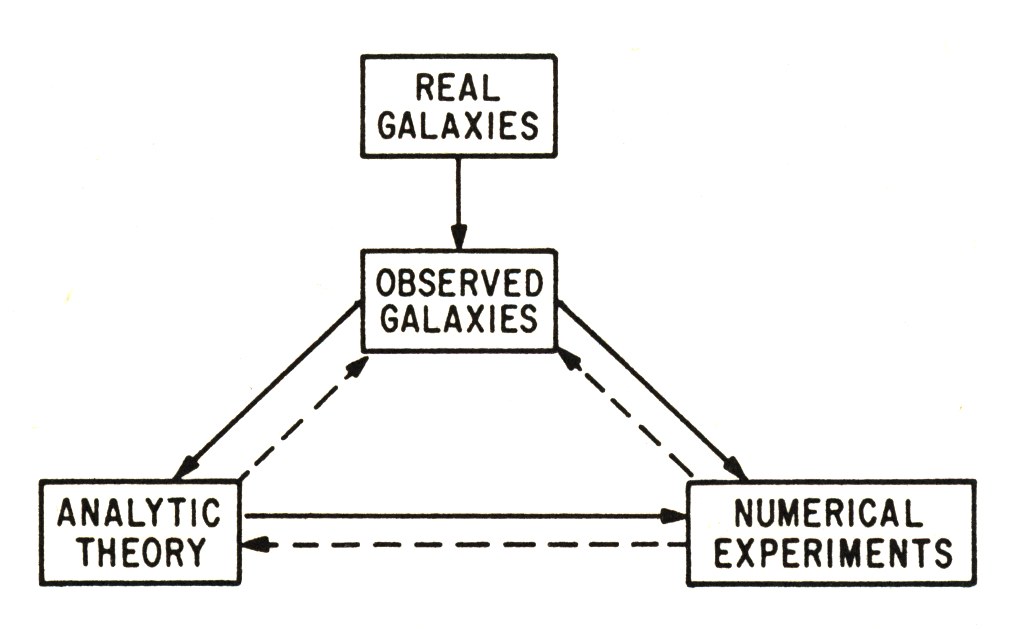} 
\caption{Logical Connections} 
\label{fg1} 
\end{center} 
\end{figure}
 
Real galaxies are the physical systems of interest; what little we know about them comes from observation, which rarely tells us about the physical proprties in any direct way. Observations are interpreted both through analytic theory and numerical experiments, and we hope that observers, analytic theoreticians, and numerical experimenters talk to each other.

The crucial question for this conference is how trajectory separation affects our understanding of the physics of galaxies (or of star clusters). Does trajectory separation compromise our treatments by analytic theory? By numerical experiments? These questions do not seem to have been clarified at this workshop.
Numerical experiments are compromised only if the numerical trajectories differ in some important way from "physical trajectories". The difference is important only if numerical trajectories visit parts of the phase space differently from physical trajectories. That would imply some integrals present in one system but not in the other. It seems unlikely, but how do we address this problem?

A purely numerical approach, such as shadowing, won't help because all trajectories being compared in shadowing are computed trajectories. We can only say how well computed trajectories compare; but if computed trajectories visit different parts of the phase space differently from physical trajectories, neither the original nor the shadow trajectory can explore the phase space freely. Both are subject to the same restrictions.

\vspace{0.2in}
\textbf{2.	Comment on Gurzadyan's Problem 2}
\vspace{0.1in}

Autocorrelations of some state functions are known \textit{not} to decay exponentially with time in stellar dynamical systems. The question is not trivial, but the result has been known for some time. This problem was studied many years ago by Chandrasekhar as he was trying to develop a "statistical stellar dynamics." He envisioned the development of a stellar system as a kind of Brownian motion, in which the transition from the state of the system at time $t$ to that at $t + dt$ is a Markov process with some by a transition matrix. Markov processes lead to exponential decays in the autocorrelations of descriptive functions. Chandrasekhar arrived at the startling conclusion that the autocorrelation in the force acting at one point within a stellar system decayed as $1/t$, rather than as exp($-\beta t$), as would be expected for a Markov process. The arguments are developed in a remarkable set of papers with von Neumann (Chandrasekhar \& von Neumann 1942, 1943) followed by a couple by Chandrasekhar alone (Chandrasekhar 1944a, 1944b). The $1/t$ result is in (1944b). The conclusion was reconfirmed a few years ago by Ed Lee (Lee 1968), using more modern language.

Chandrasekhar's result is stronger than the demonstration that the gravitational \textit{N}-body system is not an Anosov $C$-system that I mentioned earlier at this conference. My result says that the phase space does not have a simple geometric structure in which the number of expanding and contracting dimensions is constant throughout the space, which is quite a strong requirement. Chandrasekhar, on the other hand, demonstrated that the time dependence of the autocorrelation in the force acting at a given point is not exponential. His demonstration imposes less demanding requirements on the phase space.

\vspace{0.1in}
\textbf{3.	Finite Numbers of Particles}
\vspace{0.1in}

The fact that numerical experiments handle a finite number of particles has been mentioned somewhat apologetically. One should not apologize. Real galaxies and real star clusters have a finite number of particles. A limited number of particles actually makes numerical studies a more faithful representation of real galaxies than are analytic Vlasov models with an infinite number of particles.

With modes, where we expect $O$(n) modes, numerical experiments may have half a million modes, while Vlasov models have an infinite number. Neither matches a real galaxy correctly, but we can hope that the low-order modes will be similar between the two cases. This hope has been borne out in all cases where comparisons could be made, a result which gives confidence that both numerical experiments and analytic theory report those low-order modes correctly.

\vspace{0.1in}
{\footnotesize
\bf
\noindent
Chandrasekhar S., 1944a, ApJ 99, 25 \\
Chandrasekhar S., 1944b, ApJ 99, 47 \\
Chandrasekhar S., von Neumann J., 1942, ApJ 95, 489 \\
Chandrasekhar S., von Neumann J., 1943, ApJ 97, 1 \\
Lee, E. P., 1968, ApJ 151, 687
}

\vspace{0.3in}
\textbf{Avram HAYLI}
\vspace{0.1in}

My general feeling is that the very nature of the $N$-body problem will prevent from drawing precise previsions over the long term. Of course certain well behaved maps exhibit shadowing property, in the sense that near a numerically computed orbit in the phase space there exists a true orbit of different, yet unknown, initial condition than the one intended. But as we are looking for a precise description, I wonder if such a result may have any practical interest. Anyway the building of a computer code to describe the $N$-body system with a phase trajectory close to that of the theoretical one for arbitrarily long time scales will fail because of the repeated close encounters of three or more particles. Finding out a method to control  $\epsilon_c$ on the long term and finally ask $\epsilon_c$ to be 0 seems in my opinion an impossible hope.

\vspace{0.3in}
\textbf{Louis MARTINET}
\vspace{0.1in}

It seems to me that my paper "On the Permissible Percentage of Chaotic Orbits in Various Morphological Types of Galaxies" brings some elements of answer to problem 4. However, a key approach towards a general solution could consist to translate the original behaviour into terms of behaviour of geodesics on a surface with a metric defined by means of the potential of the system considered. Practically nothing exists in the literature in this frame of mind concerning applications to galaxies. In fact the realistic galactic potentials are not easily tractable in this context. However, we plan to deal with some simple cases in the near future.

\vspace{0.3in}
\textbf{Yakov PESIN (Pennsylvania State University)}
\vspace{0.1in}

As for problems 1 and 2, I hope to get some results about the decay of correlations for general systems with non-zero Lyapunov exponents. There is also that one can say about infinite measure case. I am now much interested in a couple of things including fractal geometry and fractal dimension and spatial-temporal chaos. I have done something interesting and there is a great hope to get some strong results (but for the dissipative case so far) about how one can obtain information on a dynamical system by looking at the lattice model one has while working with a computer. Another problem is to construct the gravitation theory (or something like that) on a fractal set (and I suppose that the Universe has a fractal structure). I have some tiny idea about that.

\vspace{0.3in}
\textbf{Juan Carlos MUZZIO}
\vspace{0.1in}

Concerning problems 5 and 6, on numerical codes for $N$-body systems, let us first distinguish two very different cases: a) Systems where the relaxation time is comparable to their age (e.g., open and globular clusters); b) Systems where the relaxation time is much larger than their age (e.g., galaxies). In the first case, the star-star interactions must be taken into account, so that each star in the system is represented with one body in the simulation and purely Newtonian forces are used; close encounters may demand the use of regularization techniques and, alternatively, the effect of distant masses can be averaged (as in tree codes). This problem is physically unstable: very small departures from a given initial condition will result in large departures from the final condition for the \textit{real} system, if the evolution is followed for a long enough time; therefore, to build a stable code in this case is impossible, and we can only hope that \textit{macroscopic} properties (say, half-mass radii, number of escapees, and so on) be preserved, even though \textit{microscopic} properties (individual positions and velocities) are not. In case b), instead, each body in the simulation represents millions of stars in reality and, correspondingly, the individual masses and the interparticle distances are very different in the stellar system and in the simulation. Besides, in order to get reasonably small relaxation effects in the simulation, one has to resort to codes that compute the potential through expansion in suitable basis functions, or to direct summation codes with softened force laws and rather large softening parameters. It would be desirable to obtain in the simulation trajectories similar to those of stars in the equivalent general smooth constant potential but, again, that seems an impossible task: the very fact that the relaxation effects can be reduced, but not eliminated, shows that the energies of the individual particles are not conserved, as required by motion in such a potential. Thus, we can only hope for an accurate description of macroscopic properties in this case too. In the particular case of problem 6, I believe that the new perturbation particle methods offer great promise because, using particles only for the perturbation, they greatly reduce the relaxation effects that arise from the necessarily finite number of particles that enter in the simulations.

The increase in quantity and quality of observational data (problem 8) poses an interesting challenge to theoreticians. Most, perhaps all, of us agree nowadays with the idea, pioneered by K.R. Popper, that no scientific theory can be proved to be right, it can only be proved to be wrong: no matter how many observations corroborate the theory, if a new observation contradicts it, we must change the theory (or show that the new observation was wrong!). Therefore, it is the duty of observers to find facts that could lead to the rejection of current theories, rather than to their corroboration, and no theoretician should feel bad about this. Alternatively, theoreticians should make predictions that could help the observers to disprove their theories and, in that way, aid themselves to build better theories. I think that it is not so important for a theory to be true (in the long run, all may be proved false), as it is for it to be fruitful: even if a theory turns out to be wrong, if in the process it suggested new observations that led to better theories, then it should be regarded as a very useful theory indeed. I believe that it is extremely important for theory and observation to march at a similar pace. If one advances much farther than the other, sooner or later the former will have to wait for the latter to come closer. One of my dearest Professors, the late "Don Miguel" Itzigsohn, used to remark that, while it is commonplace to note that Kepler would not have found his laws without the exquisitely precise (for their time) observations of Tycho Brahe, it is equally true that, had Tycho's observations been even more precise, Kepler could not have found his laws either, because it would have then been obvious that the planetary orbits are not exactly ellipses. The new observational material is already here, waiting to be used to devise more refined theories: the time is now, the theoreticians are us; otherwise, future technical advances will yield even better and more abundant observations that will be ever more difficult to accommodate within the present theoretical framework.

\vspace{0.3in}
\textbf{Shogo INAGAKI}
\vspace{0.1in}

It is sometimes too difficult to study the chaotic properties of real $N$-body systems because the force diverges at zero distance and the phase space extends to infinity. Therefore I would suggest to study simpler models such as I suggested in the workshop before studying real N-body systems. Though my model has some similar properties as real $N$-body systems, it has finite forces at all distances and the configuration space is compact and one-dimensional. Therefore it should be much easier to deal with.

It is also important to clarify the meaning of Gurzadyan-Savvidy time-scale.

\vspace{0.3in}
\textbf{Daniel PFENNIGER}
\vspace{0.1in}

My comments are related to Problem 4.

The problem that should be examined in any theories is to determine its fragile points. Often this task may take a long time, because a new theory is first tested against the simplest cases. Next more subtle aspects can be discovered and investigated.

Concerning the $N$-body model with respect to stellar systems, in the last decades one "subtle" point has become clear: chaos makes models fragile to perturbations. As consequence such questions should be asked: what is the scope of applicability of Liouville's theorem in chaotic stellar systems, when each trajectory or the global system is sensitive to perturbations from the rest of the Universe?

In an earlier work with Colin Norman (1990, ApJ 363, 391) I was surprised to see how chaotic orbits are also responsive to dissipative perturbations. Weak dissipative effects are \textit{amplified} by chaos. Most stellar systems do have a weak degree of dissipation and are strongly chaotic. Could a weak dissipation determines the long term state of stellar systems? In gas rich systems like spiral galaxies this is likely, but ellipticals contain also several percent of gas. Other secular dissipative effects are related to the mass loss from stars; clearly the mass lost in a stellar population after a few Gyr is not negligible.

Therefore, I doubt that "phase space volume conservation" arguments can really be applied to processes like galaxy formation by collapses or mergers of galaxies.

\vspace{0.3in}
\textbf{George S. DJORGOVSKI}
\vspace{0.1in}

I wonder if we are ready for the 10 key questions, be it these or some others. This meeting has been a success - if for no other reason, then as a very interesting experiment in the sociology of science - but we still have a long way to go! Among the mathematicians, numerical simulators, and observers, we barely even have a common language, or maybe even the common goals. Still, we have seen lots of mutual good will to learn from each other, and to understand each other. I guess we need more meetings to do our compulsory relaxation and strong mixing, and I don't mean only the drinks.

Let us cast the scene as a three-body interaction between the mathematical ergodic theory, numerical simulations (disturbingly few of which at this conference actually dealt with modeling of stellar systems!), and observations. These three things are not isomorphic, and the best they can hope is to make the life more interesting for each other, and maybe find some real-world manifestations which can be understood or plausibly explained by the theory. There are possible pitfalls all around. Mathematicians can get enamored by cleverness for its own sake, which does little good to anyone else, and it sure will not bring them any appreciation from the philistine masses of observers and suchlike lumpenproletariat. Similarly, numerical simulators can easily get lost in their fancy video games, and I have seen many tools in search of the problems, and most problems don't fit the tools. Finally, it is all too common for observers to wallow in a gross empiricism, or to pursue botanical astronomy as a hobby.

Well, speaking from my cultural bias, I think that there are some excellent problems out there, which could profit from the talents and expertise of black-belt dynamicists. Observations of real-life stellar systems pose some fascinating challenges, and I tried to point out a few of them in my written contribution to this volume. I'd rather see a real problem tackled, no matter how fuzzy, no matter how simply, than an unrealistic (and I'd say, usually sterile) toy problem. The real universe is far more interesting to some of us, than any logical construct involving grids of perfect balls and springs or similar contraptions operating without a benefit of friction, external perturbations, and similar annoyances which only exist in the real world... I think that there is a real payoff for an adventuresome dynamicist who manages to solve a problem posed by the real world, both in terms of an intellectual satisfaction, and a professional recognition.

But remember: there is no such thing as an isolated or a dissipationless stellar system out there. The universe is a chaotic mess, maybe even mathematically so.

\newpage

\begin{figure}[h]
\begin{center} 
\includegraphics[width=1.00\textwidth]{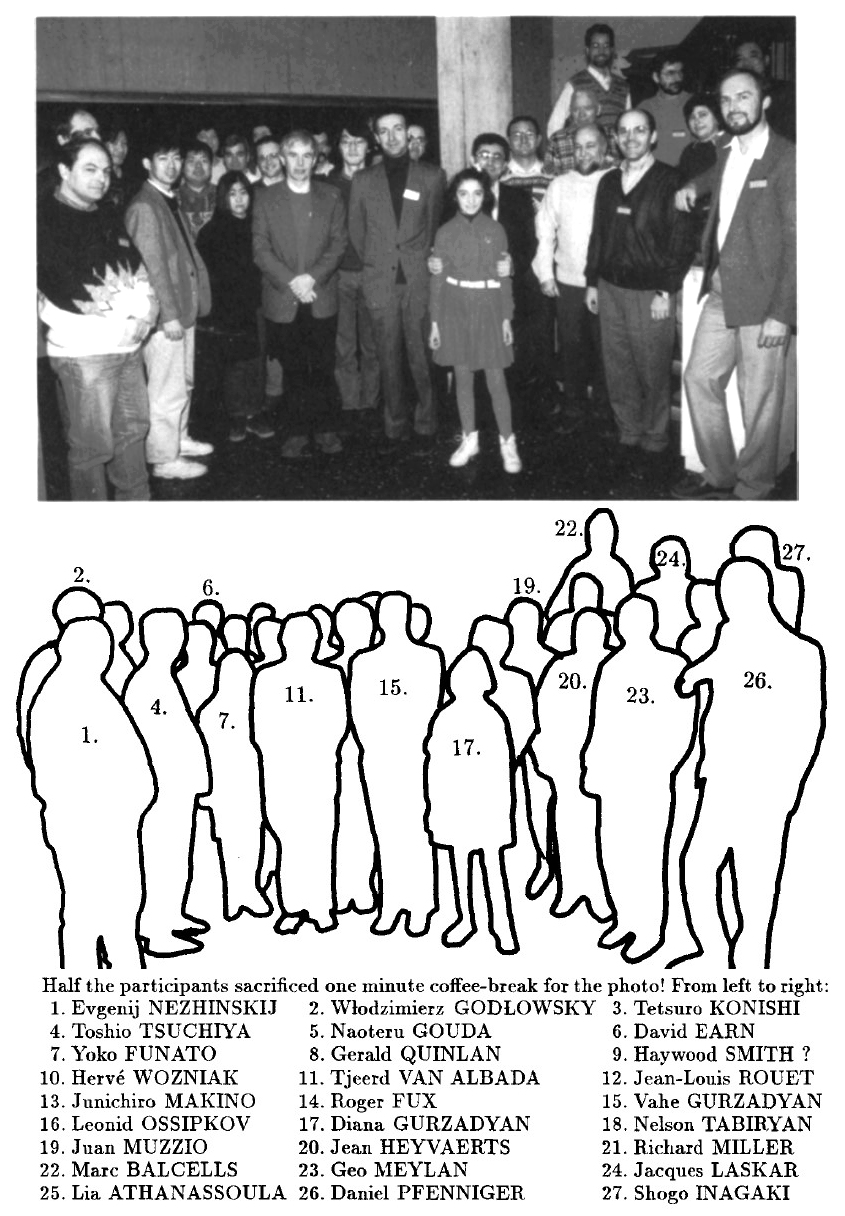} 
%\caption{Logical Connections} 
\label{fg1} 
\end{center} 
\end{figure}

\end{document}